\documentstyle[12pt]{article}
\begin{document} \begin{center}
{\Large \bf Diffusion with rearranging traps}
\vskip .5cm S. Mandal, R. Dasgupta, T. K. Ballabh, T. R. Middya and S. Tarafdar$\footnote[1]{corresponding author, email : sujata@juphys.ernet.in}$ \\Condensed Matter Physics Research Centre,\\ Physics Department, Jadavpur University\\ Calcutta -- 700032, India \end{center}

\vskip 1cm \noindent {\bf Abstract}\\ A model for diffusion on a cubic lattice with a random distribution of traps is developed. The traps are redistributed at certain time intervals. Such models are useful for describing systems showing dynamic disorder, such as ion-conducting polymers . In the present model the traps are infinite, unlike an earlier version with finite traps, this model has a percolation threshold. For the infinite trap version a simple analytical calculation is possible and the results agree qualitatively with simulation.
\vskip .25cm 
\noindent PACS nos : 05.40.+j, 66.30.Dn, 61.43.-j \vskip .5cm
\section{Introduction}Diffusion in a disordered system is a well-studied problem \cite{hav,bou}. Different approaches have been taken -- such as models which consider inter-site barriers of varying heights, or models which picture sites for diffusing particles as wells with varying depths, also combinations of barrier and well models \cite{kehr}.
Problems where the diffusing particle may encounter randomly distributed traps are also studied \cite{mur,1mur,arg}. However, one class of problems involve {\it dynamic disorder}, i.e. wells, barriers or traps whose distribution in space changes with time. In other words the location of the traps get redistributed at certain time intervals. Such models describe systems such as  ion-conducting polymers, or glasses above the glass transition temperature \cite{ratner,man}.

Models with dynamic disorder have also been studied for some time. The `dynamic bond percolation model' (DBPM)\cite{ratner,rat1} by the Northwestern University group, and its many variations study one aspect of the problem quite extensively, effective medium approaches have also been used \cite{emt,nitz}.

A different formulation of the dynamic disorder problem, considering a `well' model, rather than a `barrier' model was suggested by Bhattacharyya and Tarafdar \cite{bt}. Mandal et al \cite{man} applied model \cite{bt} to explain experimental results on the PEO-NH$_4$ClO$_4$ system \cite{ajb}.The model \cite{bt} considers a random distribution of two types of sites on a square lattice, both having finite dwell times, i.e. non-zero jump rates. It is not a `percolation' model in the sense that it never has zero diffusivity. The diffusion coefficient for different ratios of the two types of sites, and for different rearrangement times was calculated by computer simulation.

In the present work we take the rearranging lattice model with finite traps \cite{bt} to the limit where one type of site is an infinite trap. In this limit the diffusion coefficient as function of the trap concentration $c$ and rearrangement time $\tau_r$ are calculated in three-dimensions. The results are compared with the finite trap model as well as DBPM, and are supported by computer simulations.
\section{The rearranging trap model}
We consider a cubic lattice with a fraction $c$ of sites occupied by traps. A particular configuration of traps remains constant from $t=0$ to a time interval $t=\tau_r$. After that, the traps get rearranged, though the trap concentration $c$  is constant. If $N_0$ walkers started a random walk on the lattice from different sites at $t=0$, during time $\tau_r$ a certain number get trapped. The total square distance that would have been travelled by the $N_0$ walkers in the absence of traps is \begin{equation}( {r_0}^2 )_{tot} = 2.d.N_0.D_0\tau_r \end{equation} If there are traps this value gets reduced to say $ ({r_{tr}}^2 )_{tot}$. Here $D_0$ is the diffusion coefficient with no traps, and $d$ is the dimension of the system.

After the interval $\tau_r$, since traps are rearranged, previously trapped walkers may be set free and resume their walk. In the next interval from $t=\tau_r$ to $t=2\tau_r$ they again cover a total distance$( {r_{tr}}^2) _{tot}$. Fig 1 shows  how the square distance covered increases with time in our simulation described in section 3. So for a time interval $>>\tau_r$, the diffusion coefficient on a rearranging lattice is given by \begin{equation} D_{tr}=({r_{tr}}^2 )_{tot} /{2.d.N_0.\tau_r} \end{equation}
This argument is the same as in \cite{rat1}. The problem is thus reduced to calculating $( {r_{tr}}^2)_{tot}$.With traps having infinite depth (i.e. infinite dwell time), this is estimated as follows, without involving any complicated mathematics.
\subsection{Calculating the diffusion coefficient $D(c,\tau_r)$}
Suppose $N_0$ particles start a random walk at $t=0$ on a cubic lattice occupied randomly by $c$ fraction of traps. The remaining $(1-c)$ sites are `allowed' sites. If a particle encounters a trap it gets stuck until the next rearrangement time , $\langle{r_{tr}}^2\rangle$ is calculated as the mean distance travelled by each of the $N_0$ particles $( {r^2}_{tot} )/N_0$. If there were no traps $({r^2})_{tot}$ would be given by equation 1. This is the total distance travelled by all the particles in time $\tau_r$.
In presence of traps more and more particles get stuck and the total square distance travelled does not reach the value in eqn (1), but falls short by an amount $\Delta r$. \begin{equation} \Delta r = 2.d.N_0.D_0.\tau_r -  (r_{tr}^2)_{tot}  \end{equation}
knowing $\Delta r$ one can calculate $D(c,\tau_r)$ from the relation
\begin{equation} D(c,\tau_r) =  (r_{tr})^2_{tot} /{2.d.N_0.\tau_r}             \end{equation}

\subsection{Calculation of $\Delta r$ : the trapping law}
To calculate $\Delta r$ we assume a simple law for trapping.\\ We assume that the number of particles $dN$ trapped in the time interval $t$ to $t + dt$ is proportional to $N$ -- the number of particles at time $t$, $c$ -- the trap concentration and $dt$. So, \begin{equation} dN = -a.c.N.dt \end{equation} $a \sim dS(t)/dt $, where $S(t)$ is the number of distinct sites visited in time $t$ \cite{hvd}. For three or higher dimensions $$ S(t) \propto t $$ \cite{mon}, so $a$ is a constant depending on the dimension of the system but independent of $t$. We have thus the simple trapping rule \begin{equation} N=N_0.exp(-a.c.t) \end{equation}

It may be mentioned here, that problems of absorption by traps, is  well studied, and exact analytical results have been worked out for the asymptotic limit $t\rightarrow \infty$ \cite{dons,gras}. In the asymptotic limit the number of surviving particles at time t is given by \begin{equation} N = N_0.exp({-a.{c^{2/(d+2)}}.{t^{d/(d+2)}}}) \end{equation}
where the exponents of $c$ and $t$ as well as $a$ are dimension dependent. However this result is valid only when the probability of survival becomes very small $\sim 10^{-13}$ \cite{surv}. In the regime we are interested in the simple expression eq(6) agrees much better with simulation results than eq (7) as we show in fig 2 for trapping in three-dimensions. We may now proceed to calculate $D$. The particles trapped at time $t$ to $t=t + dt$ contribute an amount $\delta r$ less to $ (r_{tr}^2)_{tot} $ than they would if not trapped. \begin{equation} \delta r = 2.d.(\tau_r-t) \end{equation} Henceforth we consider the case $d=3$ only. 

Integrating this over the interval $\tau_r$ for all particles trapped during this time, we have \begin{equation} \Delta r ={ \int_{t=0}^{\tau_r}} \delta r dN = 6{\int_0}^{\tau_r} a.c.N_0.exp({-a.c.t})dt.(\tau_r-t).D_0  \end{equation}
or, \begin{equation} = 6.D_0.N_0(\tau_r +exp({-a.c.\tau_r})/{a.c} -1/{a.c}) \end{equation}
Since \begin{equation}  {(r_{tr}^2)}_{tot}  = 6.N_0.D_0\tau_r - \Delta r \end{equation} the diffusion coefficient in the presence of traps with a rearrangement time $\tau_r$ is \begin{equation} D=(D_0/ac\tau_r)(1-e^{-ac\tau_r})(1-c) \end{equation}
the factor $(1-c)$ accounts for the particles which are trapped at the outset i.e. at $t=0$.
\subsection{The limiting cases}
For $\tau_r \rightarrow 0$ that is for very rapid rearrangement \begin{equation} D/D_0 = 1-c \end{equation} as expected. For $\tau_r \rightarrow \infty $ that is when the system is effectively quenched, with no rearrangement \begin{equation} D/D_0=1 \end{equation} for $c=0$ and \begin{equation} D/D_0=0 \end{equation}for $c>0$. So there is a percolation threshold at $c=0$. Any nonzero trap concentration, however small gives $D=0$ in the limit $\tau_r \rightarrow 0$. Fig 3 shows $D$ vs $c$ for different ${\tau_r}$. The value of $a$ is taken as $a=.68$, as given by our simulation results for  the survival of random walkers in a three-dimensional lattice with traps.
\section{Simulating diffusion on the rearranging trap model}
We have simulated the random walk on a three-dimensional rearranging lattice, and calculated the diffusion coefficient for different $c$ and $\tau_r$. The algorithm is somewhat similar to the finite trap model in \cite{bt}.

We work on an effectively infinite lattice, thus avoiding finite size effects. A random walker starts to walk on a cubic lattice of unit spacing with a concentration $c$ of traps. The location of the traps are not preassigned, the walker decides whether the current site is a trap from a random number, as it goes on. However, once assigned, the trap location remains fixed for time $\tau_r$ until the next rearrangement. Each trap is of infinite depth, so the walker encountering a trap has to remain there until the next rearrangement. To incorporate this in the algorithm, as soon as the walker falls in a trap, we freeze that walk upto the next rearrangement. To ensure that a site once assigned as `allowed', is not seen as a trap on a subsequent visit, we maintain a list of all allowed sites visited during $\tau_r$, so that if these are revisited they are still assigned as `allowed'.

We average over $10^5$ walks to get $\langle {r^2} \rangle $ for different time intervals $t>>\tau_r$.$\langle {r^2} \rangle $ vs. $t$ are shown in fig 1, the curve is exactly like the results shown by \cite{rat1}. $D(c,\tau_r)$ calculated from the average slope of $\langle {r^2} \rangle $ vs. $t$ are plotted against $\tau_r$ for different $c$ and shown in fig. 4. In fig. 5 we compare the simulated value of $D$ with the value calculated from equation 12.
\section{Discussion}

Figure 5 shows that the calculated results are always lower than the value obtained from our simulations. The reason is probably that our simplified calculation ignores the distribution of distances travelled by different walkers and assumes an equal average distance covered by all the walkers. However the agreement is quite satisfactory inspite of this.

The present model though closely related to the DBPM \cite{rat1} and the finite trap model \cite{bt} shows features distinct from either of them. We discuss here the points of similarity and dissimilarity between the models.

Physically the difference between the finite trap model \cite{bt}  and the present one is only that here the traps are infinite. At this limit however, a qualitative difference enters which is the appearance of the percolation threshold. Now, in the quenched limit $\tau_r \rightarrow \infty $, when the trap positions are frozen, the diffusion coefficient is zero for nonzero $c$, so the threshold is at $c=0$. The position of the threshold is at $c=0$, whatever be the dimension of the system. So, the difference with the model \cite{bt} is that there is a threshold. On the other hand the DBPM does have a percolation threshold $p_c$, which is at $c=0$ for 1-dimension, but for higher dimensions it is at the appropriate bond percolation threshold for that dimension.

Our simulation results shown in figure 1 resemble the DBPM results for $p<p_c$. So, the trapping effect is stronger here than in DBPM. This is expected since with sites as infinite traps, a random walker must fall in it given sufficient time, even if the trap concentration is very small. In the bond model `forbidden' sites are inaccessible for the walker and may be avoided for $c<p_c$.

The finite trap model was shown to account successfully for the dynamic disorder in the polymer electrolyte $PEO-NH_4Cl_4$ \cite{man}. There the crystalline regions in the polymer are assumed to be the sites with longer dwell time, and the amorphous regions the highly conducting sites with smaller dwell time. This model is suited for systems with stronger trapping, and also systems where reacting species are diffusing in a viscous medium. For example -- formation for excimers in fluorescence experiments, where  fluorescing monomers combine with each other to form an `excited dimer' or {\it excimer} \cite{exc}. The excimer has a certain lifetime after which it may dissociate, i.e. the walker is released from the trap. However, in this case $dN \propto c^2$ rather than $c$ in equation (5) if $c$ is the monomer concentration.

The primary interest of this model is that it is possible to calculate the dynamic disorder effect in a very simple way, compared to the DBPM which requires very involved mathematics, and the results are qualitatively very similar.

\section{Acknowledgement}
We thank UGC for financial assistance. S.Mandal  and R.Dasgupta acknowledge CSIR for the award of Senior Research Fellowship and Research Associateship respectively. Authors also thank Dr. P. Nandy for useful discussions. 

\newpage
\noindent
{\bf Figure Captions}
\vskip 1.5cm
\noindent
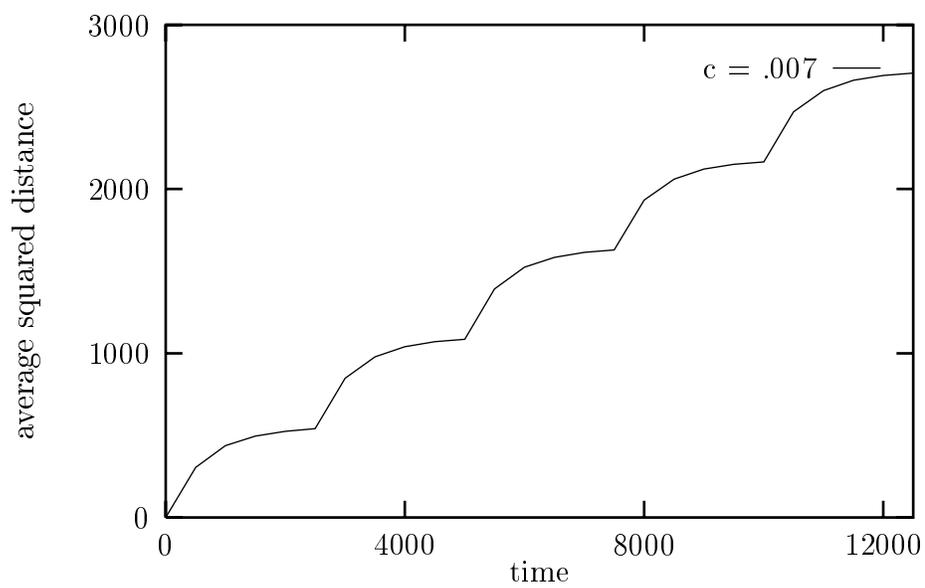
{\bf Figure 1} : Plot of average squared distance versus time for c = .007 from computer simulation.
\vskip 1cm
\noindent
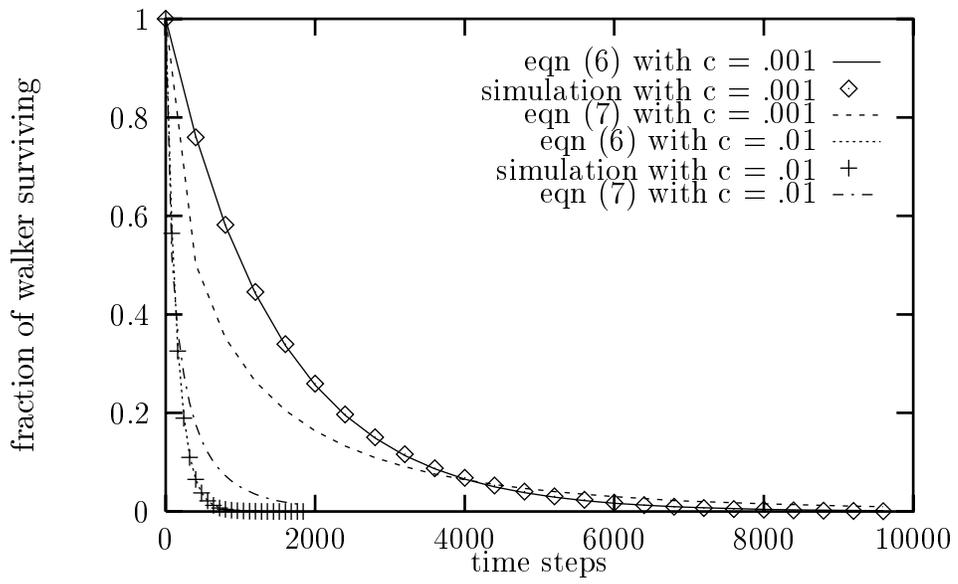
{\bf Figure 2} : Plot of fraction of walkers surviving versus time steps from simulation, equation (6) \& equation (7) for c = .001 \& c = .01
\vskip 1cm
\noindent
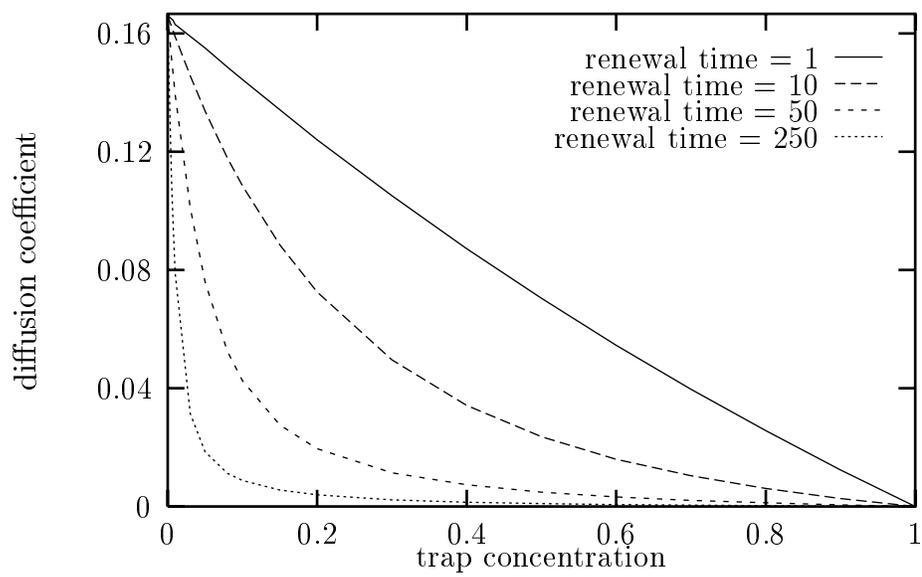
{\bf Figure 3} : Plot of calculated diffusion coefficient versus trap concentration for renewal time = 1, 10, 50, 250.
\vskip 1cm 
\noindent
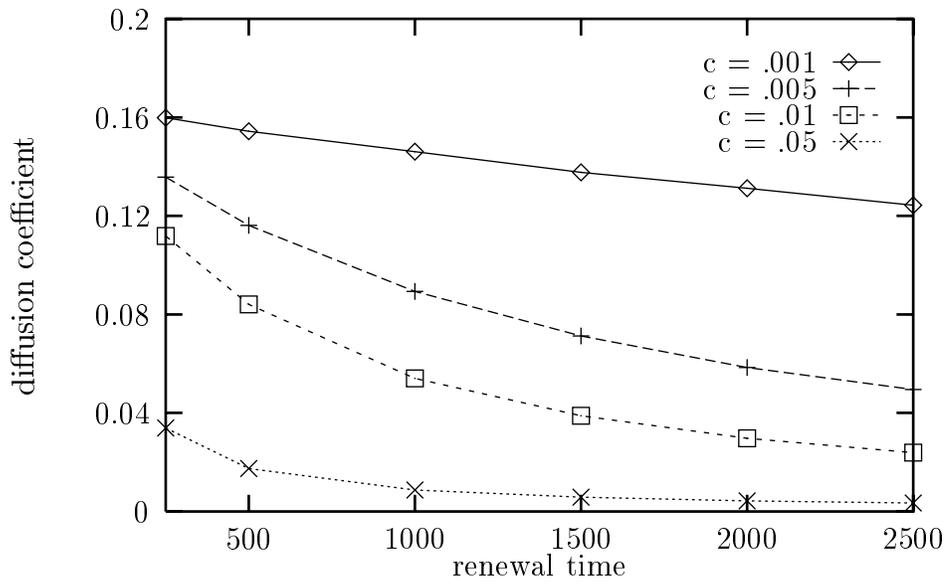
{\bf Figure 4} : Plot of diffusion coefficient versus renewal time for c = .001, c = .005, c = .01 \& c = .05 from simulation.
\vskip 1cm
\noindent
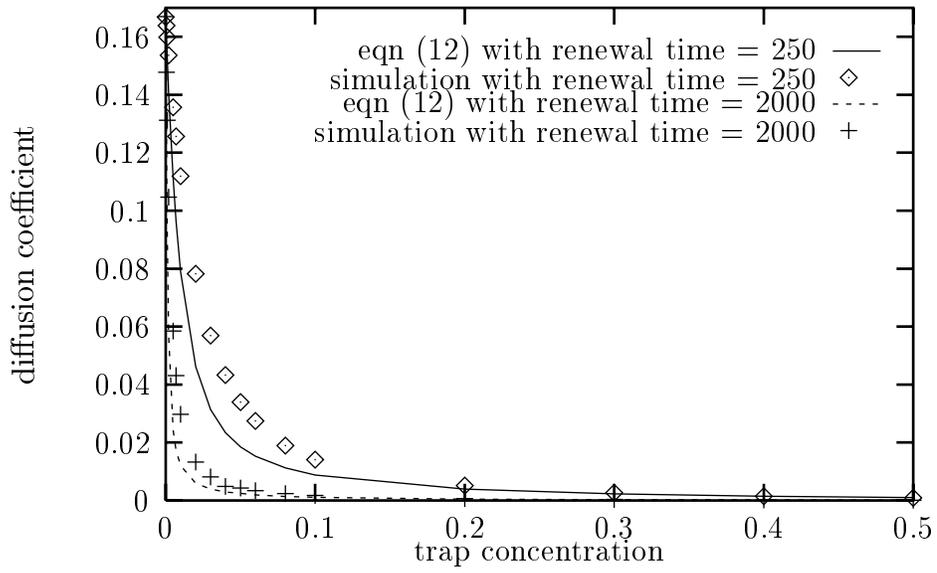
{\bf Figure 5} : Comparison of diffusion coefficient versus trap concentration  from simulation and theory [equation (12)] for renewal times = 250, 2000.


\begin{thebibliography}{99}
\bibitem{hav}S. Havlin and D. Ben Avraham, Adv. Phys., {\bf 36},(1987) 695.
\bibitem{bou}J. P. Bouchaud and A. Georges, Phys. Rep.,{\bf 195},(1990) 127.
\bibitem{kehr}K. Mussawisade, T. Wichmann and K. W. Kehr, J. Phys. Cond. Mat, {\bf 9},(1997) 1181 and references therein.
\bibitem{mur}T. Wichman, A. Giacometti, K. P. N. Murthy, Phys. rev. E, {\bf 52}, (1995) 481.
\bibitem{1mur} A. Giacometti, K. P. N. Murthy, Phys. Rev. E,{\bf 53}, (1996) 5647.
\bibitem{arg}P. Argyrakis, K. W. Kehr, J. Stat. Phys., {\bf 63}, (1991) 400.
\bibitem{ratner}A. Nitzan, M. A. Ratner, J. Phys. Chem., {\bf 98},(1994) 1765.
\bibitem{man}S. Mandal, S. Tarafdar and A. J. Bhattacharyya, Sol. State Comm.,{\bf 113} (2000) 611.
\bibitem{rat1}S. D. Druger, A. Nitzan and M.A. Ratner, J. Chem. Phys.,{\bf 79},(1983) 3133.
\bibitem{emt}A. K. Harrison and R. Zwanzig, Phys. Rev. A,{\bf 32},(1985) 1072.
\bibitem{nitz}R. Granek and A. Nitzan, J. Chem. Phys.,{\bf 90},(1989) 3784.
\bibitem{bt}A. J. Bhattacharyya, and S. Tarafdar, J. Phys. Cond. Mat.,{\bf 10},(1998) 931.
\bibitem{ajb}A. J. Bhattacharyya, T. R. Middya and S. Tarafdar, Phys. Rev. B,{\bf 60} (1999-II) 909.
\bibitem{hvd}`Fractals' by M. Daoud and H. V. Damme (pg 81) in {\it Soft Matter Physics}, ed. M. Daoud and C. E. Williams (Springer, 1995).
\bibitem{mon}E. W. Montroll and B. J. West, {\it On An Enriched Collection Of Stochastic Processes}, Studies in Statistical Mechanics VII, ed. E. W. Montroll and J. L. Lebowitz (North Holland, Amsterdam 1979).
\bibitem{dons}N. D. Donsker and S. R. S. Varadhan, Commun. Pure Appl. Math., {\bf 32},(1979) 721.
\bibitem{gras}P. Grassberger and I. Proccacia,J. Chem. Phys., {\bf 77}, (1982) 6281.
\bibitem{surv}S. Havlin, M. Dishon, J.E. Kiefer and G. H. Weiss, Phys. Rev. Lett.,{\bf 53}, (1984) 407.
\bibitem{exc}J. B. Birks, Rep. Prog. Phys.,{\bf 38}, (1975) 903.
\end{thebibliography}
\end{document}